\documentstyle[aps,psfig,multicol]{revtex}
\newcommand{\bcols}{\ifpreprintsty\else\begin{multicols}{2}\fi}
\newcommand{\ecols}{\ifpreprintsty\else\end{multicols}\fi}
\begin{document}
\draft
\title{Limits to error correction in quantum chaos}
\author{P. G. Silvestrov$^{1,2}$, H. Schomerus$^{1}$\cite{HS}, and C. W. J.
Beenakker$^{1}$}
\address{$^1$ Instituut-Lorentz, Universiteit Leiden, P.O. Box 9506, 2300 RA
Leiden, The Netherlands\\ $^2$ Budker Institute of Nuclear Physics, 630090
Novosibirsk, Russia}
\date{December 2000}
\maketitle
\begin{abstract}
We study the correction of errors that have accumulated in an entangled state
of spins as a result of unknown local variations in the Zeeman energy ($B$) and
spin-spin interaction energy ($J$). A non-degenerate code with error rate
$\kappa$ can recover the original state with high fidelity within a time
$t_{\rm R}\simeq\hbar\kappa^{1/2}/{\rm max}(B,J)$ --- independent of the number
of encoded qubits. Whether the Hamiltonian is chaotic or not does not affect
this time scale, but it does affect the complexity of the error-correcting
code.
\end{abstract}
%\narrowtext
\bcols
In classical mechanics, chaos severely limits the operation of a reversible
computer \cite{Ben85}. Any uncertainty in the initial conditions is magnified
exponentially by chaotic dynamics, rendering the outcome of the computation
unpredictable. This is why practical computational schemes are irreversible:
Dissipation suppresses chaos and makes the computation robust to errors
\cite{Lan96}. A quantum computer does not have this option; It relies on the
reversible unitary evolution of entangled quantum mechanical states, which does
not tolerate dissipation \cite{Gru99}. This invites the question
\cite{Geo99,Fla99}, what limitations quantum chaos might pose on quantum
computing.

To answer this question one needs to consider the possibilities and
restrictions of quantum error correction \cite{Ste98}. Errors can occur due to
interaction with the environment (errors of decoherence) and due to uncertainty
in the unitary evolution (unitary errors). The original state can be recovered
reliably if the errors involve at most a fraction $\kappa\lesssim 0.1$ of the
total number of qubits. The corresponding maximal time during which errors may
be allowed to accumulate (the recovery time $t_{\rm R}$) is easy to find if
different qubits are affected independently. That may be a reasonable
assumption for certain mechanisms of decoherence and also for unitary errors
resulting from an uncertain single-particle Hamiltonian. Uncertainties in the
interactions among the qubits pose a more complex problem \cite{Gea98}.

Georgeot and Shepelyansky \cite{Geo99} studied this problem for a model
Hamiltonian of $N$ interacting spins that exhibits a transition from regular
dynamics (nearly isolated spins) to chaotic dynamics (strongly coupled spins).
They concluded for the chaotic regime that $t_{\rm R}$ goes to zero $\propto
1/N$ for large $N$, but their analysis did not incorporate the optimal
error-correcting procedure. We assume a good (non-degenerate) error-correcting
code and obtain a recovery time of the order of the inverse energy uncertainty
per spin --- irrespective of the number of encoded qubits. By considering both
phase-shift and spin-flip errors we find that $t_{\rm R}$ is insensitive to
whether the Hamiltonian is chaotic or not. (The authors of Refs.\
\cite{Geo99,Fla99} arrived at the opposite conclusion that $t_{\rm R}$
increases strongly when chaos is suppressed, but they took only spin-flip
errors into account.) The absence of chaos can be used to reduce the complexity
of the code, in that a {\em classical\/} error-correcting code suffices for the
majority of the errors in the regime of regular dynamics.

The Hamiltonian $H$ under consideration describes $N$ coupled spins
$\mbox{\boldmath$\sigma$}\!_{n}=(\sigma_{n}^{x},\sigma_{n}^{y},\sigma_{n}^{z})$
on a lattice in a magnetic field ${\bf B}_{n}=(B_{n}^{x},B_{n}^{y},B_{n}^{z})$,
\begin{equation}
H=\sum_{n}{\bf B}_{n}\cdot\mbox{\boldmath$\sigma$}\!_{n}+\sum_{n\neq
m}\mbox{\boldmath$\sigma$}\!_{n}\cdot {\bf
J}_{nm}\cdot\mbox{\boldmath$\sigma$}\!_{m}.\label{Hdef}
\end{equation}
A spin $n$ interacts with $d$ neigboring spins $m$ via the matrix ${\bf
J}_{nm}$. The spin could be a nuclear spin or the spin of an electron confined
to a quantum dot, in the context of solid-state based proposals for quantum
computing \cite{Pri98,Kan98,Los98}. More generally, the spin could be a
representation of a two-level system (for example, in the context of the
ion-trap quantum computer \cite{Cir95}). We assume that ${\bf B}_{n}$ and ${\bf
J}_{nm}$ fluctuate independently from site to site, with zero mean and variance
$\overline{|{\bf B}_{n}|^{2}}=B^{2}$ and
$\sum_{\alpha\beta}\overline{(J_{nm}^{\alpha\beta})^{2}}=J^{2}$ (provided $m$
is one of the $d$ neighbors of $n$, otherwise ${\bf J}_{nm}=0$). We denote by
$U=(B^{2}+2dJ^{2})^{1/2}$ the root-mean-square energy uncertainty per spin.

A state $\psi_{0}$ evolves in time according to $\psi(t)=e^{-iHt}\psi_{0}$
(setting $\hbar\equiv 1$). We assume that we do not know the parameters of the
Hamiltonian, and use quantum error correction to recover $\psi_{0}$ from
$\psi(t)$ \cite{note1}. Let $\psi_{0}$ lie in the code space of a
non-degenerate error-correcting code \cite{Ste98}. The code space is a $2^{M}$
dimensional subspace of the full $2^{N}$ dimensional Hilbert space, such that
\begin{equation}
\langle\psi_{0}|\sigma^{\alpha_{1}}_{n_{1}}\sigma^{\alpha_{2}}_{n_{2}}
\ldots\sigma^{\alpha_{k}}_{n_{k}}|\psi'_{0}\rangle=0,\;\;1\leq k\leq
2K,\label{codedef}
\end{equation}
for any two (possibly identical) states $\psi_{0},\psi'_{0}$ in the code space
and any product of up to $2K$ Pauli matrices $\sigma_{n}^{\alpha}$ (acting on
different spins $n_{1},n_{2},\ldots$). The number $M$ is the number of qubits
encoded in $N$ spins. The number $K$ is the number of errors that the code can
correct, where the application of $\sigma_{n}^{x},\sigma_{n}^{y}$, or
$\sigma_{n}^{z}$ to any of the $N$ spins counts as one error. The ratio
$M/N=\rho$ is the bit rate of the code and the ratio $K/N=\kappa$ the error
rate.

Error correction is successful if $\psi(t)$ lies in the error space of
$\psi_{0}$, which is the subspace spanned by the state $\psi_{0}$ and the
states derived from $\psi_{0}$ by making up to $K$ errors. The operator ${\cal
P}$ projects onto the error space. Explicitly: ${\cal P}=\sum_{p=0}^{K}{\cal
P}_{p}$, with
\begin{equation}
{\cal P}_{p}=\sum_{\{n,\alpha\}}\frac{1}{p!}\sigma_{n_{1}}^{\alpha_{1}}
\ldots\sigma_{n_{p}}^{\alpha_{p}}|\psi_{0}\rangle
\langle\psi_{0}|\sigma_{n_{1}}^{\alpha_{1}} \ldots\sigma_{n_{p}}^{\alpha_{p}}.
\label{Ppdef}
\end{equation}
The symbol $\sum_{\{n,\alpha\}}$ indicates a summation over the $n_{i}$'s and
$\alpha_{i}$'s, with the restriction that the indices $n_{1},n_{2}\ldots$
should be all distinct. (The indices $\alpha_{1},\alpha_{2}\ldots$ need not be
distinct.) The norm
\begin{equation}
F(t)=|{\cal P}\psi(t)|^{2}=\langle\psi_{0}|e^{iHt}{\cal
P}e^{-iHt}|\psi_{0}\rangle\label{Ftdef0}
\end{equation}
of the projected state is the probability of successful error correction after
a time $t$. It is the ``fidelity'' of the recovered state \cite{Ste98}. The
recovery time $t_{\rm R}$ can be defined as the time at which the fidelity has
dropped from 1 to 1/2.

We assume that the error-correcting code is ``good'', meaning that $\rho$ and
$\kappa$ tend to a non-zero value as $N\rightarrow\infty$. Good quantum-error
correcting codes exist, but their construction for large $N$ is a complex
problem \cite{Cal96,Ste96,Got96,Eke96,Cal97,Kni00}. Our strategy will be to
derive a lower bound to $F$ and $t_{\rm R}$ that does not use any properties of
the code beyond the non-degeneracy condition (\ref{codedef}), so that we can
avoid an explicit construction. An alternative approach would be to abandon the
requirement of a ``good'' code, and keep the number $M$ of encoded qubits fixed
as the total number of spins $N$ goes to infinity. One can then use the
technique of concatenation \cite{Ste98} to construct codes that are safe for a
large number of errors at the expense of a vanishingly small bit rate $\rho$.
(See Ref.\ \cite{Gea00} for such a calculation in the case $M=1$.)

Our first step is to decompose the evolution operator
$e^{iHt}=\sum_{k=0}^{N}X_{k}$ into operators $X_{k}$ that create $k$ errors.
For $k\ll N$ and $t\ll 1/U$ we may approximate
\begin{eqnarray}
X_{k}&=&X_{0}\sum_{q=0,2,4}^{k}\sum_{\{n,\alpha\}}
\frac{(it)^{k-q/2}}{(k-q)!(q/2)!} 
\sigma_{n_{1}}^{\alpha_{1}}\ldots \sigma_{n_{k}}^{\alpha_{k}}
\nonumber\\
&&\mbox{}\times  J_{n_{1}n_{2}}^{\alpha_{1}\alpha_{2}}\ldots
J_{n_{q-1}n_{q}}^{\alpha_{q-1}\alpha_{q}} B_{n_{q+1}}^{\alpha_{q+1}}\ldots
B_{n_{k}}^{\alpha_{k}} ,\label{XkX0relation}\\
X_{0}&=&\exp\biggl[-\case{1}{2}t^{2}\sum_{n}|{\bf B}_{n}|^{2}-t^{2}\sum_{n\neq
m}\sum_{\alpha,\beta}(J_{nm}^{\alpha\beta})^{2}\biggr]. \label{X0result}
\end{eqnarray}
The approximation consists in neglecting terms in the exponent of order
$k(Ut)^{2}$ and $N(Ut)^{4}$, relative to the terms retained of order
$N(Ut)^{2}$. We may write $X_{0}\approx\exp[-\case{1}{2}N(Ut)^{2}]$, neglecting
fluctuations in the exponent that are smaller by a factor $1/\sqrt N$.

We next substitute the decomposition of $e^{iHt}$ in error operators into the
fidelity (\ref{Ftdef0}),
\begin{equation}
F(t)=\sum_{p=0}^{K}\sum_{k,k'=0}^{N}\langle X_{k} {\cal
P}_{p}X_{k'}^{\ast}\rangle, \label{Ftdef}
\end{equation}
where we have abbreviated $\langle\cdots\rangle=
\langle\psi_{0}|\cdots|\psi_{0}\rangle$. To simplify this expression, we take
the average over the random variations in the $B_{n}$'s and $J_{nm}$'s. (We
will show later that statistical fluctuations around the average are
insignificant.) Only the terms with $k=k'$ contribute to the average. The terms
with $p+k\leq 2K$ can be simplified further, since they contain at most $2K$
Pauli matrices. In view of Eq.\ (\ref{codedef}), these expectation values
vanish unless the product of Pauli matrices reduces to a c-number, which
requires $p=k$. Hence the average fidelity can be written as
$\bar{F}=F_{1}+F_{2}$, with
\begin{eqnarray}
F_{1}&=&\sum_{p=0}^{K}\sum_{\{n,\alpha\}}\frac{1}{p!}\overline{|\langle X_{p}
\sigma_{n_{1}}^{\alpha_{1}} \ldots\sigma_{n_{p}}^{\alpha_{p}}\rangle|^{2}},
\label{F1def}\\
F_{2}&=&\sum_{p=0}^{K}\sum_{k=2K+1-p}^{N}\sum_{\{n,\alpha\}}\frac{1}{p!}
\overline{|\langle X_{k} \sigma_{n_{1}}^{\alpha_{1}}
\ldots\sigma_{n_{p}}^{\alpha_{p}}\rangle|^{2}}. \label{F2def}
\end{eqnarray}

\begin{figure}
\centerline{\psfig{figure=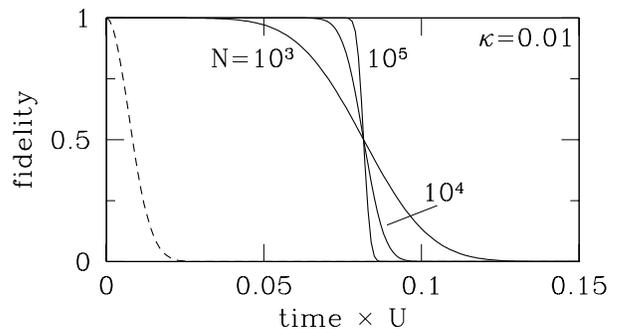,width=8cm}\medskip}
\caption[]
{Solid curves: time dependence of the lower bound $F_{1}$ to the
ensemble-averaged fidelity, calculated from Eq.\ (\protect\ref{F1largeK}) for
error rate $\kappa=0.01$ and three values of $N$. (We took $B^{2}=2dJ^{2}$, so
that the root-mean-squared energy uncertainty per spin $U$ is equally divided
between Zeeman energy and interaction energy.) The dashed curve shows (for
$N=10^{4}$) the squared overlap $X_{0}^{2}=\exp[-N(Ut)^{2}]$ between initial
and final state. In all these curves the number $M$ of encoded qubits is a
fixed fraction $\rho$ of the total number of spins $N$.
\label{fidelity}}
\end{figure}

The expectation values in $F_{1}$ are evaluated by substituting Eq.\
(\ref{XkX0relation}) and extracting the terms that reduce to a c-number,
\begin{equation}
F_{1}=e^{-N(Ut)^{2}}\sum_{p=0}^{K}\sum_{q=0,2,4}^{p}
\frac{(NB^{2}t^{2})^{p-q}}{(p-q)!}\frac{(2NdJ^{2}t^{2})^{q/2}}{(q/2)!}.
\label{F1result}
\end{equation}
For $K\gg 1$ we may approximate $e^{-x}x^k/k!\approx(2\pi
x)^{-1/2}\exp[-(k-x)^{2}/2x]$ and replace the sums in Eq.\ (\ref{F1result}) by
integrals. The result is
\begin{mathletters}
\label{F1largeK}
\begin{eqnarray}
&&F_{1}=\case{1}{2}+\case{1}{2}{\rm erf}[(t_{\rm R}-t)/\Delta
t],\label{F1largeKa}\\
&&t_{\rm R}=\sqrt{\frac{\kappa}{B^{2}+4dJ^{2}}},\;\; \Delta
t=\sqrt{\frac{1}{2N}}
\frac{\sqrt{B^{2}+8dJ^{2}}}{B^{2}+4dJ^{2}},\label{F1largeKb}
\end{eqnarray}
\end{mathletters}%
with ${\rm erf}(x)=2\pi^{-1/2}\int_{0}^{x}e^{-y^{2}}dy$ the error function.
Corrections to $t_{\rm R}$ and $\Delta t$ arising from the approximations made
in Eqs.\ (\ref{XkX0relation}) and (\ref{X0result}) are smaller by a factor
$\kappa$. The expectation values in $F_{2}$ depend specifically on $\psi_{0}$,
hence on the way in which $M$ qubits are encoded in $N$ spins. Since $F_{2}\geq
0$ we have a lower bound $\bar{F}\geq F_{1}$ on the fidelity that is {\em code
independent} within the class of non-degenerate error-correcting codes.

For $N\rightarrow\infty$ the time dependence of Eq.\ (\ref{F1largeK})
approaches the step function $\theta(t_{\rm R}-t)$. The threshold $t_{\rm R}$
is independent of $N$, while the width $\Delta t$ of the transition vanishes as
$N^{-1/2}$ (solid curves in Fig.\ \ref{fidelity}). These are results for the
ensemble-averaged fidelity, but since the variance is bounded by $0\leq {\rm
var}\,F\leq \bar{F}(1-\bar{F})$ the fluctuations are insignificant except in
the narrow transition region. The step-function behavior of the fidelity also
implies that the positive code-dependent term $F_{2}$ that we have not included
in Fig.\ \ref{fidelity} satisfies $\lim_{N\rightarrow\infty}F_{2}\rightarrow 0$
for $t<t_{\rm R}$ (since $F_{1}+F_{2}\leq 1$ and $F_{1}\rightarrow 1$ for
$t<t_{\rm R}$). Any code-dependence of the fidelity can therefore only appear
for times greater than $t_{\rm R}$.

The independence of the recovery time on the number $M$ of encoded qubits
disagrees with Refs.\ \cite{Geo99,Fla99}. These authors calculated the squared
overlap $|\langle\psi_{0}|\psi(t)\rangle|^{2}\approx X_{0}^{2}$ of the
time-dependent state with the original state, and argued that the original
state would be effectively lost once this overlap is $\ll 1$. However, the
original state can be recovered even when this overlap has become exponentially
small, if a good error-correcting code is used (compare dashed and solid curves
in Fig. \ref{fidelity}). The recovery time is increased by a factor
$\sqrt{\kappa M}$, with an overhead of $1/\rho$ spins per encoded qubit.

We find that $t_{\rm R}$ at a given $U$ is insensitive to the relative
magnitude of $B$ and $J$, and hence {\em insensitive to whether the Hamiltonian
is chaotic or not.} This conclusion may seem surprising in view of the fact
that the eigenstates are completely different in the chaotic and regular
regimes \cite{Geo99}: For $J<B/N$ the eigenstates of the total Hamiltonian $H$
are a superposition of a small number of eigenstates of the non-interacting
part $\sum_{n}{\bf B}_{n}\cdot\mbox{\boldmath$\sigma$}\!_{n}$. This number
(known as the participation ratio) increases with increasing $J$, and when
$J\approx B$ it becomes of the same order as the dimension $2^{N}$ of the
entire Hilbert space. (See Ref.\ \cite{Sil98} for a description of the onset of
quantum chaos in systems with random two-body interactions.) As we will now
discuss, the reason that a small participation ratio does not improve the
fidelity is that it counts spin-flip errors but not phase-shift errors. For the
same reason, suppression of chaos does help to reduce the complexity of the
error-correcting code.

The three Pauli matrices correspond to three types of errors: spin flips
($\sigma^{x}$), phase shifts ($\sigma^{z}$), and a combination of the two
($\sigma^{y}=i\sigma^{x}\sigma^{z}$). The complexity of the code is reduced
substantially if there is only one type of error to correct. (One can then use
a code for classical bits, such as the Hamming code \cite{Ste98}.) Suppose that
we seek to suppress spin-flip errors, of either type $\sigma^{x}$ or
$\sigma^{y}$. To this end we impose on the spins a known uniform magnetic field
in the $z$-direction, with Zeeman energy $B_{0}$ that is large compared to the
magnitude $U$ of the random energy variations. The new Hamiltonian is
$H+H_{0}$, with $H$ given by Eq.\ (\ref{Hdef}) and
$H_{0}=B_{0}\sum_{n}\sigma_{n}^{z}$. Since $B_{0}$ is known we can undo the
evolution of a state due to $H_{0}$ by applying the operator
$e^{iH_{0}t}=\prod_{n}(\cos B_{0}t+i\sigma_{n}^{z}\sin B_{0}t)$. Any remaining
deviation of $\psi(t)$ from $\psi_{0}$ has to be dealt with by the
error-correcting code, with projection operator ${\cal P}$. The fidelity of the
corrected state is $F(t)=|{\cal P}G(t)\psi_{0}|^{2}$, where the evolution
operator $G$ is defined by
\begin{equation}
G(t)=e^{iH_{0}t}e^{-i(H+H_{0})t}={\cal
T}\exp\biggl(-i\int_{0}^{t}H(t')dt'\biggr).\label{Gdef}
\end{equation}
(The notation ${\cal T}$ indicates time ordering of the operators
$H(t)=e^{iH_{0}t}He^{-iH_{0}t}$.)

For $B_{0}t\gg 1$ we may replace $H(t)$ by its time average over the interval
$(t,t+1/B_{0})$. The terms containing a single $\sigma^{x}$ or $\sigma^{y}$
average out to zero and we are left with
\begin{eqnarray}
&&G(t)=e^{-it(H_{\parallel}+H_{\perp})},\;\;
H_{\parallel}=\sum_{n}B_{n}^{z}\sigma_{n}^{z}+\sum_{n\neq
m}J_{nm}^{zz}\sigma_{n}^{z}\sigma_{m}^{z},\nonumber\\
&&H_{\perp}=\sum_{n\neq m}\case{1}{2}(J_{nm}^{xx}+J_{nm}^{yy})
(\sigma_{n}^{x}\sigma_{m}^{x}+\sigma_{n}^{y}\sigma_{m}^{y}).
\label{Gttimeaverage}
\end{eqnarray}
(We have assumed $J_{nm}^{xy}=J_{nm}^{yx}$, so that the mixed terms
$\sigma_{n}^{x}\sigma_{m}^{y}$ cancel.) The time dependence of the fidelity is
again given by Eq.\ (\ref{F1largeK}), with $B^{2}=\overline{(B_{n}^{z})^{2}}$
and $J^{2}=\overline{(J_{nm}^{zz})^{2}}+ \frac{1}{4}\overline{(J_{nm}^{xx}
+J_{nm}^{yy})^{2}}$. The recovery time $t_{\rm R}$ depends only weakly on the
ratio $J/B$. The relative number of phase-shift and spin-flip errors, however,
depends strongly on this ratio. Indeed, if one would use a code that corrects
up to $K_{\parallel}$ errors from $\sigma^{z}$ and up to $K_{\perp}$ errors
from $\sigma^{x}$ or $\sigma^{y}$, then the maximal $t_{\rm R}$ (at fixed
$K_{\parallel}+K_{\perp}$) is reached for $K_{\perp}/K_{\parallel}=
4dJ^{2}/B^{2}$. For $J\ll B$ one has $K_{\perp}\ll K_{\parallel}$, so that a
classical error-correcting code suffices for the majority of errors.

Before concluding we briefly consider the case that the parameters ${\bf
B}_{n}$ and ${\bf J}_{nm}$ in the Hamiltonian are not only unknown but also
time dependent. The result (\ref{F1result}) still holds if we replace
$(Bt)^{2}$ by the correlator
$b(t)=\int_{0}^{t}dt'\int_{0}^{t}dt''\,\overline{{\bf B}_{n}(t')\cdot{\bf
B}_{n}(t'')}$, and similarly replace $(Jt)^{2}$ by
$j(t)=\int_{0}^{t}dt'\int_{0}^{t}dt''\,
\sum_{\alpha\beta}\overline{J_{nm}^{\alpha\beta}(t')
J_{nm}^{\alpha\beta}(t'')}$. For a short-time correlation one has
$b(t)=b_{0}|t|$, $j(t)=j_{0}|t|$. This leads for $K\gg 1$ to the fidelity
\begin{mathletters}
\label{F1largeKt}
\begin{eqnarray}
&&F_{1}=\case{1}{2}+\case{1}{2}{\rm erf}[(t_{\rm R}-t)/\Delta
t],\label{F1largeKta}\\
&&t_{\rm R}=\frac{\kappa}{b_{0}+4dj_{0}},\;\; \Delta t=\sqrt{\frac{2\kappa}{N}}
\frac{\sqrt{b_{0}+8dj_{0}}}{(b_{0}+4dj_{0})^{3/2}}.\label{F1largeKtb}
\end{eqnarray}
\end{mathletters}%
The recovery time now depends linearly on the error rate $\kappa$, but it
remains $N$-independent. The next step towards fault-tolerant computing would
be to include in the Hamiltonian a part with a known time dependence,
representing the logical gates. We leave that for a future investigation.

In conclusion, we have derived a code-independent lower bound for the fidelity
$F$ of a state that has been recovered after a unitary evolution for a time $t$
in an unknown random magnetic field and spin-spin interaction. For a large
system the transition from $F=1$ to $F=0$ occurs abruptly at a time $t_{\rm R}$
that is independent of the total number of spins $N$ and the number of encoded
qubits $M$. The magnitude of $t_{\rm R}$ is set by the inverse energy
uncertainty per spin, regardless of whether the spins are nearly isolated or
strongly coupled. The suppression of chaos that occurs when the spins are
decoupled does not improve the fidelity, because of the persistence of
phase-shift errors. Spin-flip errors can be suppressed, and this helps to
reduce the complexity of the error-correcting code.

In this work we have concentrated on the recovery from unitary errors. One
might question whether suppression of quantum chaos improves the fidelity for
recovery from errors of decoherence, in particular in view of the
``hypersensitivity to perturbation'' observed in computer simulations of
systems with a chaotic dynamics \cite{Sch96,Son00}. This question presents
itself as an interesting topic for future research.

We thank P. Zoller for a valuable discussion. This work was supported by the
Dutch Science Foundation NWO/FOM. PGS acknowledges the support of the RFBR
grant number 98--02--17905.

\ecols

\begin{references}
\bibitem{note1} In special cases one can use a spin-echo technique, instead of
quantum error correction, to invert the unitary evolution. This works for the
Hamiltonian of Refs.\ \protect\cite{Geo99,Fla99}, which has the special form
$H=\sum_{n}B_{n}\sigma_{n}^{z}+\sum_{\mu\nu}
J_{\mu\nu}\sigma_{\mu}^{x}\sigma_{\nu}^{x}$, 
the spins $\mu$ and $\nu$ belonging to two disjunct
sublattices. The spin echo is prepared by the unitary operator
$S=\prod_{\mu\nu}\sigma_{\mu}^{x}\sigma_{\nu}^{y}$, which satisfies
$SHS=-H$. The state $S\psi_{0}$ is allowed to evolve under the action
of $H$ for a time interval $t$, after which $S$ is applied a second
time. After another time $t$ the original $\psi_{0}$ is recovered,
since $e^{-iHt}Se^{-iHt}S=1$.
\end{references}
\end{document}